\documentclass{llncs}
\usepackage{fancyvrb}
\usepackage{amsmath}

\DeclareMathOperator{\DTFT}{DTFT}
\DeclareMathOperator{\FRDER}{DF}
\DeclareMathOperator{\FFT}{FFT}

\usepackage{makeidx}  % allows for indexgeneration
\usepackage{graphicx}

\begin{document}

\DefineVerbatimEnvironment{sourcecode}{Verbatim}{fontsize=\small}

\newcommand{\slfrac}[2]{\left.#1\middle/#2\right.}

\frontmatter          % for the preliminaries
\pagestyle{headings}  % switches on printing of running heads

\mainmatter              % start of the contributions
\title{A multi-GPU Programming Library for Real-Time Applications}
\titlerunning{Multi-GPU Real-Time Library}  % abbreviated title (for running head)
%                                     also used for the TOC unless
%                                     \toctitle is used
%
\author{Sebastian Schaetz \inst{1} \and Martin Uecker\inst{2}}
\authorrunning{Sebastian Schaetz et al.} % abbreviated author list (for running head)
%
%%%% list of authors for the TOC (use if author list has to be modified)
\tocauthor{Sebastian Schaetz, Martin Uecker}
\institute{BiomedNMR Forschungs GmbH at the \\
Max Planck Institute for biophysical Chemistry, Goettingen \\
\email{sschaet@gwdg.de}
\and
Department of Electrical Engineering and Computer Sciences \\
University of California, Berkeley\\ 
\email{uecker@eecs.berkeley.edu}
}

\maketitle              % typeset the title of the contribution

\begin{abstract}

We present MGPU, a C++ programming library targeted at single-node multi-GPU systems. Such systems combine disproportionate floating point performance with high data locality and are thus well suited to implement real-time algorithms. We describe the library design, programming interface and implementation details in light of this specific problem domain. The core concepts of this work are a novel kind of container abstraction and MPI-like communication methods for intra-system communication. We further demonstrate how MGPU is used as a framework for porting existing GPU libraries to multi-device architectures. Putting our library to the test, we accelerate an iterative non-linear image reconstruction algorithm for real-time magnetic resonance imaging using multiple GPUs. We achieve a speed-up of about 1.7 using 2 GPUs and reach a final speed-up of 2.1 with 4 GPUs. These promising results lead us to conclude that multi-GPU systems are a viable solution for real-time MRI reconstruction as well as signal-processing applications in general.

\end{abstract}

\section{Introduction}

Within the last five years general-purpose computation on graphics hardware has become increasingly attractive among industry and academia. The combination of convenient programming tools and libraries with the striking performance-to-dollar and performance-to-watt ratio makes graphics processing units the default solution to many data-parallel problems. Several such devices can be used in a cluster configuration with multiple nodes connected via a local area network. These systems lend themselves well to solve large high performance computing problems such as scientific simulations, as the cluster can be sized to fit the problem. Cluster implementations generally exhibit good weak scaling, i.e. they perform efficiently if the problem size increases with the number of processing units.

Real-time signal-processing problems have different requirements. Their problem size is fixed and usually defined by physical constraints. In addition, calculation results must be available before an operational deadline expires. The processing delay of a real-time algorithm must be bounded as it runs in synchronism with the data acquisition process. If such real-time applications are very compute intensive and require the floating point performance of more than one GPU, single-node desktop systems with multiple compute devices are preferable over multi-node clusters. Single-node systems exhibit higher memory bandwidth and lower latency compared to clusters, resulting in better data locality. This matches the fixed problem size of real-time applications and the requirement for strong scaling. Despite these advantages, we are not aware of any programming tools or libraries that explicitly target desktop multi-GPU systems.

In this work we present a programming library for multi-GPU systems called MGPU that supports the development of strong scaling applications. We describe the core concepts, the interface, and important implementation details of the library. Performance is evaluated with a number of micro-benchmarks. We further document our efforts of speeding up an iterative non-linear image reconstruction algorithm for real-time magnetic resonance imaging (MRI) - a prime example for computationally demanding digital signal-processing algorithms.

\section{MGPU}

MGPU is a combination of a C++ template header library and a small static link library and depends on a number of Boost C++ libraries \cite{dawes2009boost} including Test, Thread, Bind and Function. It is tested on both Linux and Windows platforms and can be built using a standard C++ compiler. MGPU is implemented as a layer on top of existing GPU computing frameworks and numerical libraries, designed for single-GPU systems, and combines them under a coherent interface. It is modelled after the C++ Standard Template Library.

The modular design of MGPU supports different frameworks such as CUDA and OpenCL. The current version of MGPU focuses on the CUDA backend as it exposes hardware features not yet supported by OpenCL. MGPU applications must be compiled for each specific target system. The build process detects various performance relevant architecture features such as the number of devices present and the capabilities of each device. This allows MGPU to enable optimized versions of functions such as peer-to-peer for inter-GPU communication which is much faster than transfers staged through the host.

Since MGPU is designed to support real-time signal-processing applications it forgoes advanced automated parallelization methods employed in modern high performance computing frameworks. Instead MGPU allows full control over the hardware at all times and carefully employs convenient abstractions such as well-known MPI-based communication functions and segmented containers. These segmented containers facilitate the implementation of hardware-aware algorithms, a prerequisite for good efficiency on distributed memory systems. In addition, MGPU does not limit access to lower level GPU computing frameworks, giving developers full access to performance relevant hardware features that might not yet be supported by MGPU.

\subsection{Runtime Environment}

MGPU contains a single-threaded and multi-threaded runtime environment to interact with all GPUs in the compute node. The multi-threaded runtime creates a thread for each GPU. The single-threaded version handles each GPU from within one thread by switching GPU contexts. The runtime is initialized by instantiating an \verb+environment+ object. The following code snippet shows how the number of devices can be selected at runtime.

\vspace{-0.5ex}
\begin{sourcecode}  
  environment e;                            // use all devices
  environment e(dev_group::from_to(0, 2));  // use device 0 and 1
\end{sourcecode}
\vspace{-0.5ex}

With the default constructor, all devices present in the system are used for computation. Specifying a \verb+dev_group+ limits the number of devices that are available for computation to a subset.

\subsection{Memory Management}

GPU based architectures are distributed memory systems: main memory is often arranged in a non-uniform manner and each compute device has a separate memory block. A basic abstraction the library employs is the device vector, a container that manages data on a single GPU. It represents the device equivalent of a vector in main memory such as \verb+std::vector+. On instantiation, it allocates the requested amount of data on the device that is associated with the calling thread. Algorithms can interface with the container through iterators. The container allows access to a raw pointer \verb+T*+ as well as a device pointer \verb+dev_ptr<T>+.

\vspace{-0.37cm}
\begin{figure}[h]
\centering
\includegraphics[width=60mm]{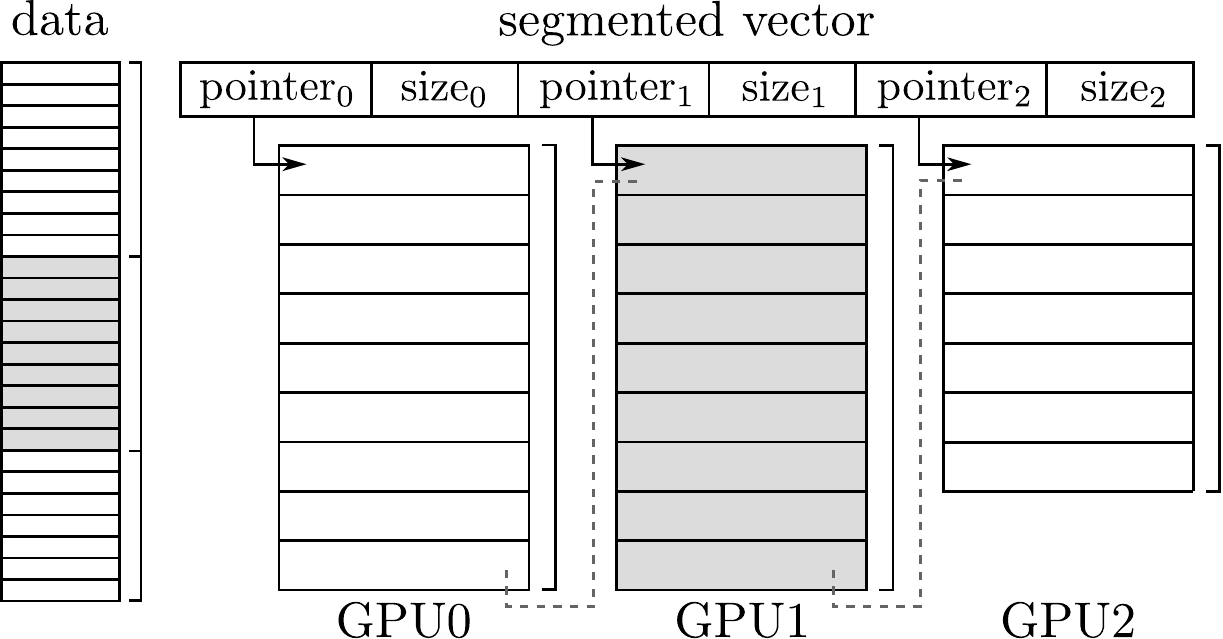}
\caption{Data stored across 3 GPUs using a segmented vector container: a vector of tuples for pointer and size represents each local device vector}
\label{fig:segdevvector}
\end{figure}
\vspace{-0.37cm}

To manage data across multiple devices, MGPU offers an implementation of the segmented container concept. It is based on the work by Austern on segmented iterators \cite{austern2000segmented}. An elementary implementation is the segmented vector that can be modelled as a vector of local vectors but is a large vector that is split in segments and distributed across the local memory of all GPUs. Figure \ref{fig:segdevvector} illustrates this for 3 GPUs. The container segments data automatically, depending on the number of compute devices used by the runtime. The way data is split across GPUs can be controlled during construction of the segmented vector. Natural and block-wise splitting as well as cloning and 2D overlapped splitting are possible. The container implicitly holds information about the location of each memory segment. This location awareness of segmented vectors makes them the major building block for implementing algorithms that scale across multiple compute devices and exploit segmentation. MGPU communication methods, linear algebra functions as well as the fast Fourier transform are aware of the data segmentation through this hierarchical abstraction. By modifying the number of devices used by the runtime environment, an algorithm can be scaled up to multiple devices.

\subsection{Data Transfer}

\begin{figure}[h]
\vspace{-0.25cm}
\centering
\includegraphics[width=80mm]{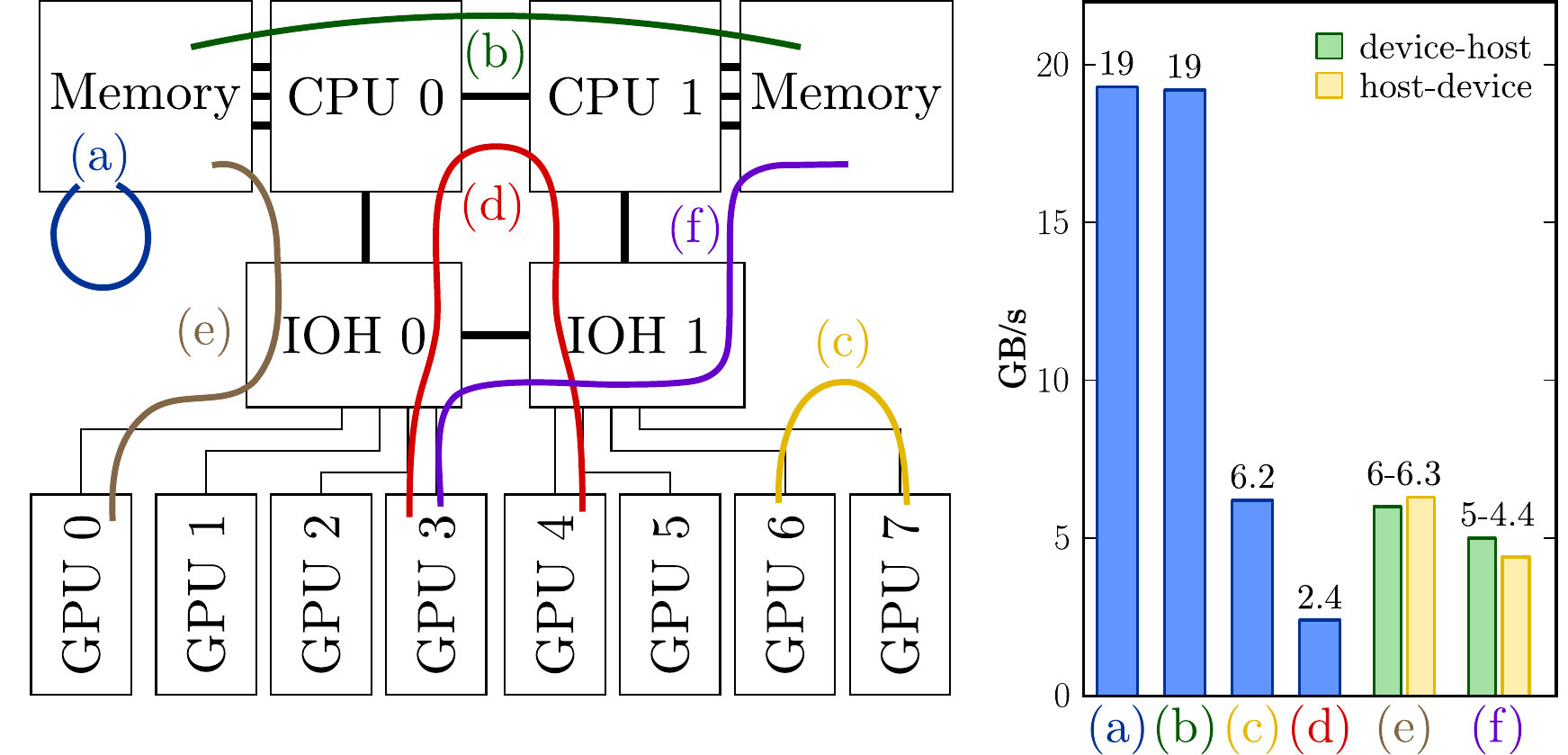}
\caption{Octo-GPU system showing various memory transfer paths and their throughput. Due to the non-uniform memory architecture, there is a difference between transfer paths (a) and (b). And since the I/O hubs (IOH) on this mainboard only support memory-mapped I/O (MMIO) between peers, GPUs connected to IOH 0 can not directly communicate with GPUs connected to IOH 1. Memory transfers between these GPUs have to be staged through main memory which accounts for the difference in (c) and (d).}
\vspace{-0.25cm}
\label{fig:tyan}
\end{figure}

Today's commercial off-the-shelf multi-GPU systems can contain up to 8 or more compute-devices. Figure \ref{fig:tyan} shows the block diagram of a state of the art Tyan FT72-B7015 computer equipped with 8 GPUs. The various memory transfer paths that are possible in such a system are highlighted. Host memory performance was measured using a NUMA version of the STEAM benchmark \cite{Bergstrom2011} and GPU memory throughput was measured using modified CUDA SDK examples.

\begin{figure}[h]
\vspace{-0.37cm}
\centering
\includegraphics[width=90mm]{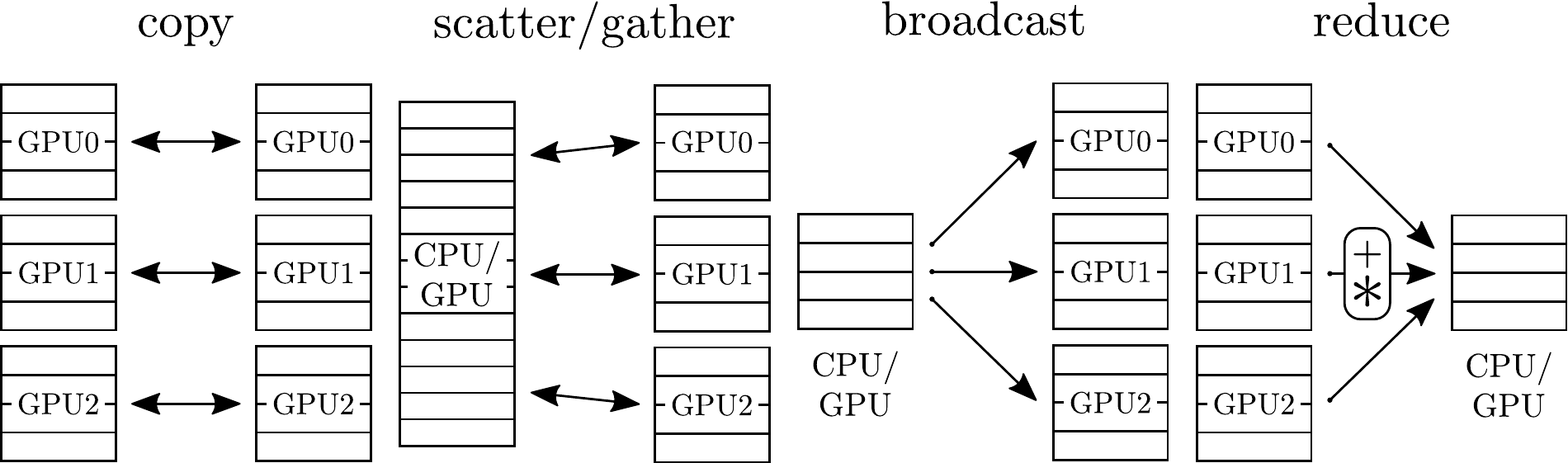}
\caption{Segmented data transfer primitives supported by MGPU: copying a segmented vector to another segmented vector, scattering and gathering a local vector from CPU or GPU to a segmented vector, broadcasting a local vector to a segmented vector and reducing a segmented vector to a local vector using an operation.}
\label{fig:transfers}
\vspace{-0.37cm}
\end{figure}

Not only are there multiple possible memory transfer paths on such devices, with segmented containers there are also various modes to transfer data. MGPU implements a subset of the MPI standard communication routines. Figure \ref{fig:transfers} shows communication primitives involving segmented containers that the MGPU library implements. 

%The primitives are united in three function signatures:
%
%\vspace{-0.5ex}
%\begin{sourcecode} 
%  copy(source_range, target_iterator);  
%  broadcast(source_range, target_iterator);
%  reduce<T, operator<T> >(source_range, target_iterator); 
%\end{sourcecode}
%\vspace{-0.5ex}

\subsection{Libraries}

MGPU also acts as a framework to use single-GPU libraries on multi-GPU systems and consolidates them under a coherent interface. Existing algorithms are extended to interoperate with segmented containers resulting in hierarchical implementations that are aware of the memory segmentation. Currently MGPU supports the CUDA FFT and the CUDA BLAS library. The interfaces are part of the modular backend architecture and can thus be ex. This will enable us to support APPML through the same interface in the future.

\vspace{-0.5ex}
\begin{sourcecode} 
  seg_dev_vector<complex<float> > d(x*y*batch, x*y);   // segment data
  fft<complex<float>, complex<float> > f(x, y, batch); // fft handle
  f.forward(d, d);                                     // distributed fft
\end{sourcecode}
\vspace{-0.5ex}

The above code snippet shows how several fast Fourier transforms can be calculated in parallel. As the input data is segmented and distributed across multiple devices, a significant speed-up can be achieved for such batched transforms. Individual FFTs can currently not be split across devices.

\subsection{Kernel Invocation and Synchronization}

Kernels can be called through the \verb+invoke+ family of functions. A kernel can be launched on all devices or just on a subset. Segmented containers are forwarded as device ranges referencing only local memory. If the entire segmented vector must be known by the kernel for peer-to-peer access, a pass-through type is provided to directly forward the segmented vector. The following listing shows various options of this mechanism:

\vspace{-0.5ex}
\begin{sourcecode} 
  invoke_kernel(kernel_caller, par1, par2, dev_rank);
  invoke_kernel_all(kernel_caller, par3, par4);
\end{sourcecode}
\vspace{-0.5ex}

The first function call invokes a kernel caller in the device context of \verb+dev_rank+. The second call launches the kernel for each device in the \verb+environment+. The \verb+kernel_caller+ function is a stub provided by the user that configures the kernel call with the proper grid- and block-size or number of work-groups and work-group size and finally calls the kernel. 

MGPU is by default asynchronous. Synchronizing separate operations and multiple compute-devices in a system is done through a family of barrier and fence functions provided by MGPU. Calling for example
\vspace{-0.5ex}
\begin{sourcecode} 
  barrier_fence();
\end{sourcecode}
\vspace{-0.5ex}

blocks all devices until all devices finished pending operations. Synchronization functions are implemented using condition variables and GPU driver synchronization mechanisms for device-local operations and incur a respective overhead.

\subsection{Evaluation}

A number of micro-benchmarks help to measure the performance of core functions of MGPU and are used to assess the benefits of using a multi-GPU system in combination with our library. Type and size of the test input data is chosen explicitly with typical real-time applications in mind.

\begin{figure*}
\vspace{-0.37cm}
\centering
\includegraphics[width=\linewidth]{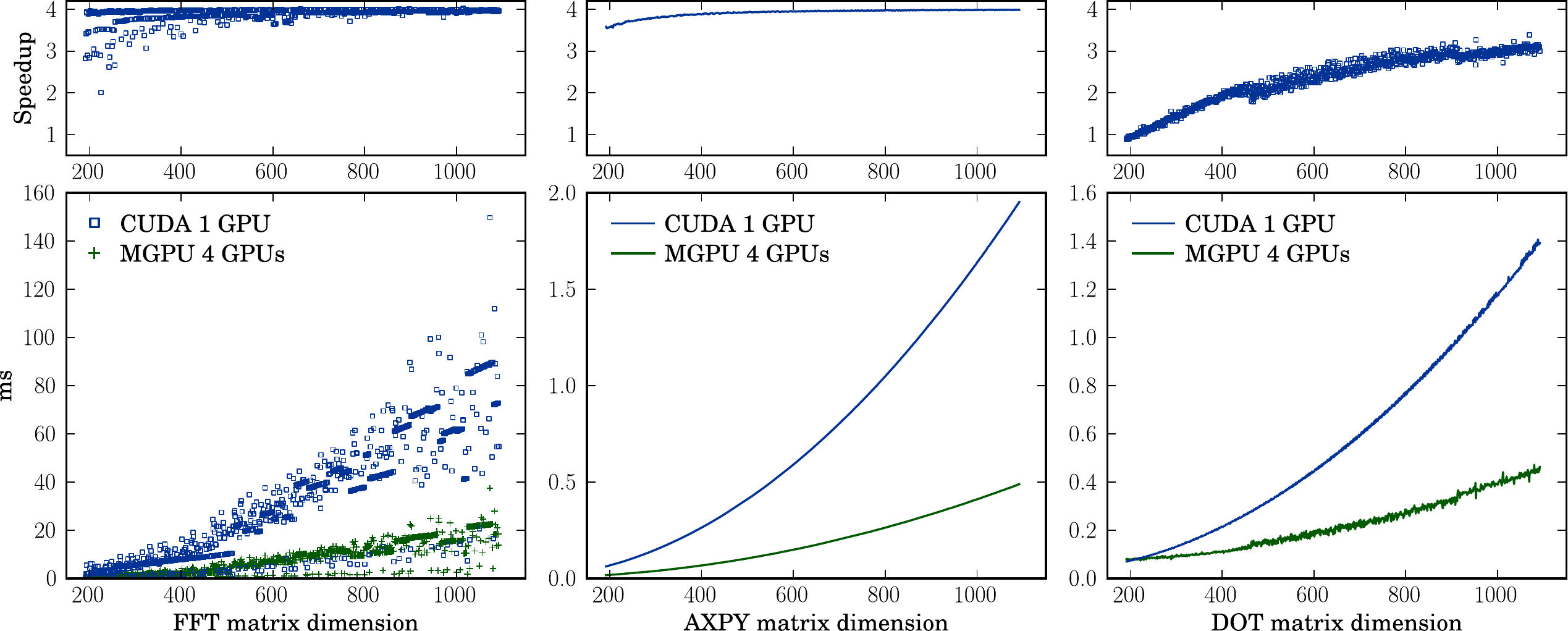}
\caption{Algorithm performance comparison of the fast Fourier transform, and basic linear algebra subprograms $a*X+Y$ as well as $A\cdot B$. Input data are 12 complex square matrices of single precision floating point values. The MGPU implementation internally calls CUDA functions.}
\label{fig:fftbench}
\vspace{-0.37cm}
\end{figure*}

The first benchmark compares the runtime of three common algorithms. Figure \ref{fig:fftbench} shows that both FFT and $a*X+Y$ operations scale well, especially for larger matrix sizes. The variance in the FFT performance is caused by the CUFFT implementation. The measured time shows the combined performance of the forward and inverse Fourier transform. The $A\cdot B$ operation does not exhibit strong scaling and an efficiency of $\frac{3}{4}$ can be achieved only for larger data sets. This is due to the reduction step in the operation that can not distribured across devices efficiently and requires and additional inter-device reduction step for the final result.

\begin{figure}[h]
\vspace{-0.37cm}
\centering
\includegraphics[width=0.9\linewidth]{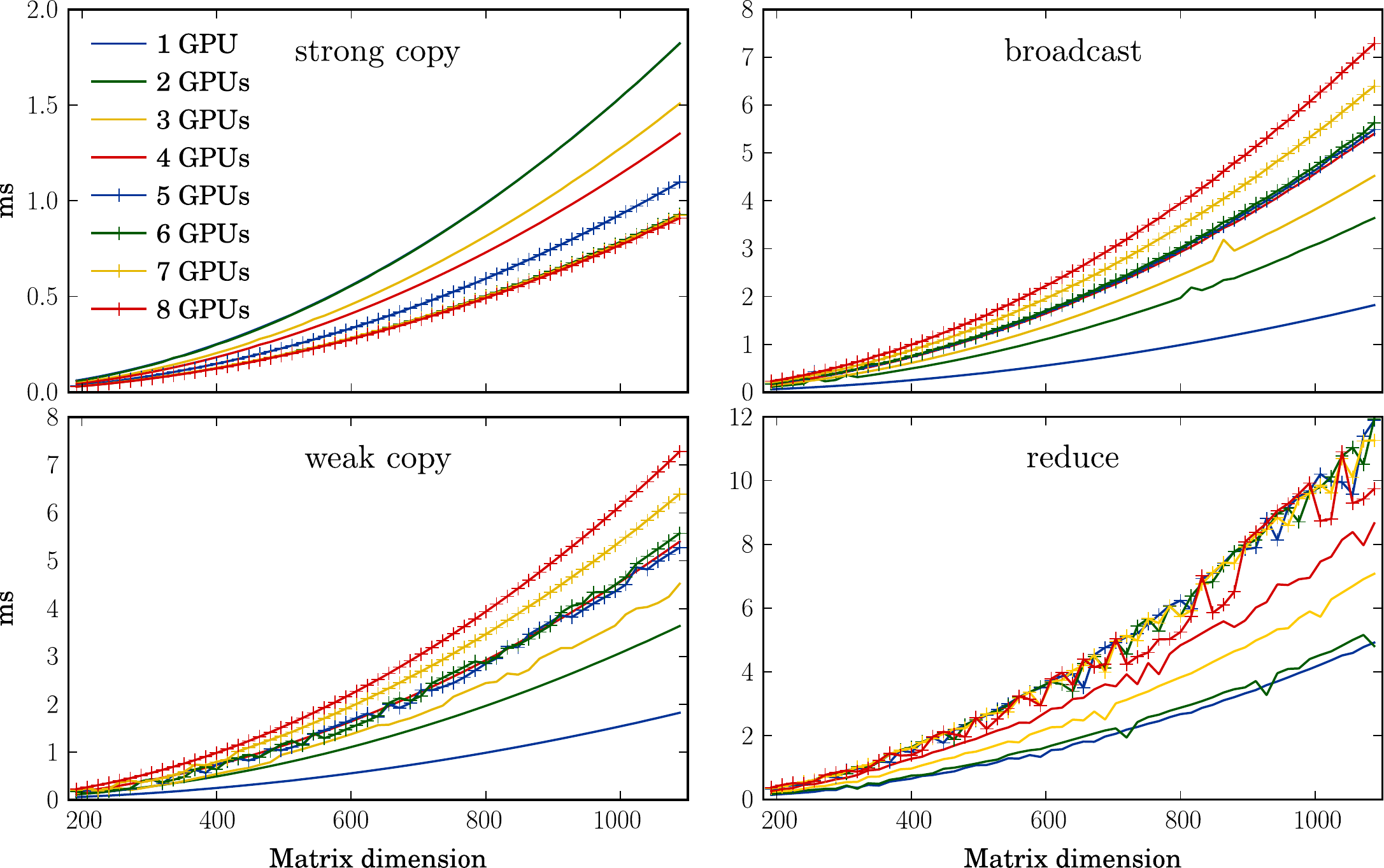}
\caption{MGPU data transfer primitives; data used for these test are squared complex single floating point matrices. Strong copy keeps the number of matrices constant with varying number of GPUs. Weak copy increases the number of matrices with the number of GPUs. The broadcast function copies one matrix to all GPUs and the reduce function merges one matrix per GPU through summation and the final result is transferred to host memory.}
\label{fig:reducebench}
\vspace{-0.37cm}
\end{figure}

Figure \ref{fig:reducebench} illustrates the performance of MGPU host to device and device to host data transfer primitives. The strong copy test, where the amount of data is constant, shows that data can be copied faster to multiple GPUs. The weak copy test behaves similarly to the broadcast test: transfer time increases with more GPUs and more data but also shows the same behaviour as the strong copy test: data can be copied more efficiently to multiple GPUs. If more GPUs are involved in the memory transfer, multiple PCI Express pathways can be utilized resulting in higher bandwidth. Reduction is done using peer-to-peer communication. 1 GPU of each PCIe domain performs a reduction through peer-to-peer data access and directly transfers the result to CPU memory. If GPUs attached to different I/O hubs are involved, peer-to-peer access is not possible between all devices and a final reduction has to be calculated by the host. For our test system this is the case if more than 4 GPUs participate. From 1 to 2 GPUs, there is negligible overhead but with increasing peer-to-peer transfer the parallel efficiency decreases.

After assessing the relative performance of individual operations using artificial benchmarks, an image reconstruction algorithm with existing single-GPU implementation is modified using MGPU to support multiple compute-devices.

\section{MRI Image Reconstruction}

Recently, iterative image reconstruction algorithms for MRI have made unprecedented temporal resolution at high image quality attainable. This is made possible by reducing the data necessary for image reconstruction to a minimum (see for example \cite{tsao2003k,block2007tv,lustig2007sparse,uecker2008image,jung2009k}). These new methods pave the way for new insights into human physiology and are a valuable new tool for scientists and radiologist to study dynamic processes such as the beating heart, joint movement and speech production.

Due to the algorithms' iterative nature, they are orders of magnitude more compute-intensive than traditional methods and reach the limits of modern high-performance computer systems. Fast image reconstruction, however, is key to facilitate adoption of these advantageous methods in a clinical setting. In such an environment, images should be reconstructed without a perceivable delay to not compromise clinical workflow and maintain patient throughput.

In the following, we document our efforts of porting an existing single-GPU implementation of the nonlinear inverse reconstruction algorithm~\cite{uecker2008image,uecker2010gpu} to a multi-GPU system using the MGPU library. Using this particular algorithm, it is possible to acquire real-time MRI movies up to a temporal resolution of 20ms~\cite{uecker2010realtime}. This achievable frame-rate widens the gap between acquisition and reconstruction time and is the primary motivation to look into new ways of accelerating the reconstruction beyond the possibilities of the existing single-GPU implementation.
 
\subsection{Reconstruction Problem and Algorithm}

During data acquisition multiple radio frequency coils $J$ positioned around the subject measure the MRI signal. Each coil possesses a unique spatial sensitivity map $c_j$. The signal equation for an image $\rho$ is:

\begin{equation}
y_j(t) = \int_{\Omega}d \vec{x} \rho( \vec{x} ) c_{j}( \vec{x} ) e^{-i\vec{k}(t)\vec{x}} 
\end{equation}

where $\vec{k}$ describes the trajectory in $k$ space and $y_j$ the signal received in coil $j$. The algorithm interprets this as an ill-conditioned nonlinear inverse problem $Fx=y$. The operator $F$ maps the unknown image $\rho$ to the acquired $k$ space positions using the sensitivities $c_j$. On a discretized rectangular Cartesian grid, the operator $F$ can be written as

\begin{equation}
	F = P_k \DTFT  M_\Omega C W^{-1} 
\end{equation}

where $P_k$ is a projection onto the measured sample positions, $\DTFT$ is the multidimensional discrete-time (here: space) Fourier transform and $M_\Omega$ is a mask that restricts the reconstructed image to the area $\Omega$. The non-linear operator $C$ multiplies image and coil sensitivities. $W$ is a weighted discrete Fourier transform applied to the coil sensitivities, which represents an assumption about the smoothness of $c_j$ that is included in the equation.

Equation $Fx=y$ is then solved using the iteratively regularized Gauss-Newton method \cite{bakushinsk2004}. In each iteration $x_{n+1}$ is estimated from the current result $x_n$ by solving the regularized equation 

{\setlength\arraycolsep{0.01em}
\begin{eqnarray}
&(& \FRDER^{\text{H}}_{x_n} \FRDER_{x_n} +  \alpha_n I ) (x_{n+1} - x_n) \nonumber \\ 
 && =  \FRDER^{\text{H}}_{x_n} (y-Fx_n) - \alpha_n (x_n - x_{\text{ref}})\;\text{.}
\label{eq:gaussnewton}
\end{eqnarray}
}

with the conjugate gradient algorithm. The regularization parameter is $\alpha_n$. After an initial interpolation of the data to the grid which is performed as a pre-processing step on the CPU, all
further operations can be performed on the grid. The operation $\DTFT^{-1} P_k \DTFT$ embedded in the left-hand side
of the equation can be understood as a convolution with the point spread function and is implemented
by applying two Fourier transforms.

%Conceptually the algorithm minimizes the difference between measured data $y$, and the data $Fx_n$ that the current estimate $x_n$ for image and coil sensitivities would produce.

\subsection{Single- and multi-GPU Implementation}

The data used in this work is acquired using a $3$T MRI system (Tim Trio, Siemens Healthcare, Erlangen, Germany) that is capable of recording up to $32$ data channels simultaneously. For real-time applications, spatial image resolution usually varies between $1.5 \times 1.5\; mm^2$ and $2 \times 2\; mm^2$ which yields a matrix size of $192$ to $384$ depending on the field of view. The grid size is then doubled to implement the non-periodic convolution with the point spread function and to achieve a high accuracy when initially interpolating the measured data onto the Cartesian grid. Computation is carried out in complex single precision floating point format. A principal component analysis preprocessing step is applied before reconstruction to compress the 32 channels to $8-12$ \cite{Huang2007compression}. The multi-GPU target system for this application is a Tyan FT72-B7015 computer equipped with two Intel X5650 six-core processors, 96GB main memory and 8 GeForce GTX 580 compute-devices with 1.5GB memory each.

The algorithm consists of a number of Fourier transform calculations applied separately to each channel, point-wise matrix operations involving one fixed matrix and the separate channel matrices as well as scalar products of all data. The original implementation utilizes the parallel computation capabilities of a single GPU to individually speed up each operation. The CUDA FFT library batched-mode is used to calculate the Fourier transforms of all channel matrices. Custom CUDA kernels handle the point-wise operations and the CUDA BLAS library is used to calculate the scalar products. Single pixels are mapped to GPU threads, image and coil sensitivities are calculated at the same time.

Successive frames in a movie are calculated subsequently, as each $x_n$ depends on the result of the previous frame $x_{\text{ref}}$. This temporal regularization makes it impossible to use a straight-forward pipeline-approach to parallelize the reconstruction of movies on a multi-GPU system. The MGPU implementation of the algorithm instead distributes the coil sensitivity maps $c_j$ across all GPUs $G$ and the image $\rho$ is split into one $\rho_{g}$ per GPU with $\rho = \sum^{G}\rho_{g}$. This summation amounts to a block-wise all-reduce operation, since all GPUs require $\rho$. An alternative decomposition is not feasible due to the FFT that operates on an entire matrix.

MGPU simplifies the process of extending the existing implementation to support multiple compute-devices. Existing containers can be replaced with segmented vectors. Only kernel interfaces have to be modified to accept local ranges instead of containers but kernel bodies can be reused and called through \verb+invoke_kernel+ functions. The FFT and BLAS library of MGPU exhibits a custom C++ interface and all calls to those libraries have to be changed. In addition, since MGPU operations are asynchronous, synchronization points have to be chosen carefully as to ensure completion of dependent operations.

Table \ref{tbl:opbreak} gives a breakdown of the three operators from equation \ref{eq:gaussnewton} and one entry that includes all additional conjugate gradient operations.

\renewcommand*\arraystretch{1.1}
\renewcommand{\tabcolsep}{6pt}

\vspace{-0.37cm}
\begin{table}

\caption{Algorithm operator breakdown showing the number of Fourier transforms, element-wise operations, channel summations, scalar products and communication steps in each operator.}
\centering
    \begin{tabular}{|l|c|c|c|c|c|}
        \hline
        ~                       & $\FFT$ & $AB$ & $\sum c_j$ & $A \cdot B$ & $\sum \rho_{n_g}$ \\ \hline
        $F$                     & 2      & 4    & ~                     & 1        & ~                  \\ 
        $\FRDER$                & 2      & 5    & ~                     & ~        & ~                  \\ 
        $\FRDER^{\text{H}}$     & 2      & 4    & 1                     & ~        & 1                  \\ 
        CG                      & ~      & 6    & ~                     & 2        & ~                  \\
        \hline
    \end{tabular}
    
\label{tbl:opbreak}

\end{table}
\vspace{-0.37cm}

While $F$ is only required once per Newton step, $\FRDER$ and $\FRDER^{\text{H}}$ are applied in each conjugate gradient iteration with the consequence of continual inter-GPU data transfer. $\sum \rho_{g}$ is implemented using the peer-to-peer communication capabilities of the GeForce GTX 580 GPUs. This restricts the number of parallel GPUs used for image reconstruction to 4 on our particular system because direct peer-to-peer communication is only possible between those devices. The following listing shows the peer-to-peer communication kernel for 4 GPUs.

\vspace{-0.5ex}
\begin{sourcecode} 
__global__ void kern_all_red_p2p_2d(cfloat * dst, cfloat* src0, 
  cfloat* src1, cfloat* src2, cfloat* src3, int dim, int off)
{
  int i = off + blockIdx.x * dim + threadIdx.x;
  dst[i] = src3[i] + src2[i] + src1[i] + src0[i];
}
\end{sourcecode}
\vspace{-0.5ex}

Each GPU runs this kernel at the same time. To speed up this operation further, the kernel only transfers a 2D section of $\rho_{g}$. This is possible because a mask $M_\Omega$ is applied immediately after the summation. A double-buffering solution is chosen to avoid local overwrites of data that is not yet transferred to neighbouring devices. Therefore, destination and local source pointers reference distributed memory blocks.

\subsection{Results}

The goal is an implementation of a non-linear iterative image reconstruction algorithm that is capable of calculating images without perceivable delay. While images can be measured at up to 50 Hz, clinically relevant protocols are envisioned to encompass frame rates of 5-30 Hz. The existing single-GPU implementation is capable of reconstructing about $3-4.5$ frames per second. This version however is not taken into account in this comparison because of MGPU-unrelated code optimization. Instead we benchmark the algorithm with varying number of GPUs since MGPU permits a seamless adjustment of the number of GPUs used for reconstruction.

\begin{figure}[ht]
\vspace{-0.37cm}
\begin{minipage}[t]{0.495\linewidth}
\centering
\includegraphics[width=\linewidth]{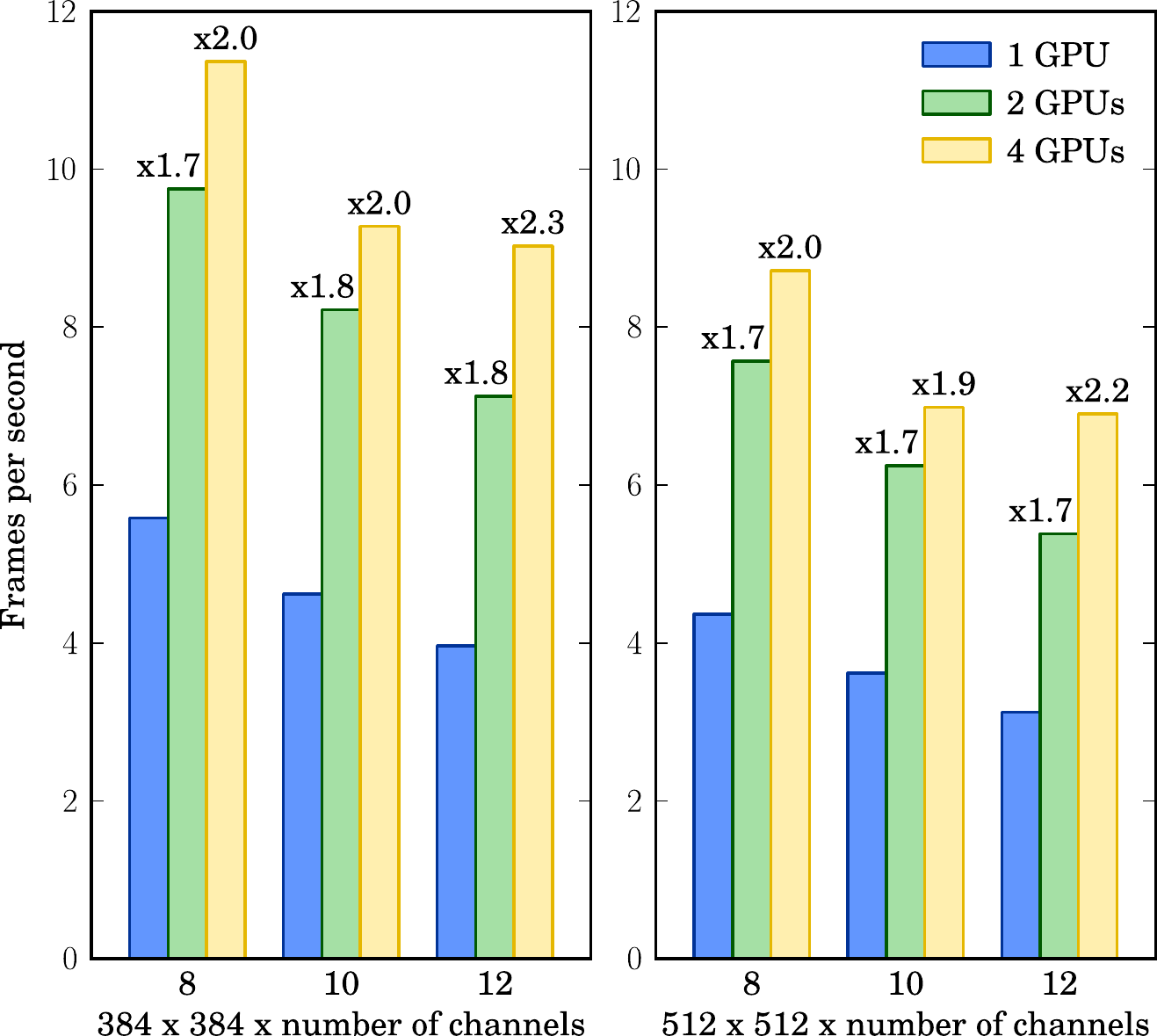}
\caption{Performance comparison with varying number of GPUs, channels, and size}
\label{fig:speedup}
\end{minipage}
\hspace{0.1cm}
\begin{minipage}[t]{0.465\linewidth}
\centering
\includegraphics[width=\linewidth]{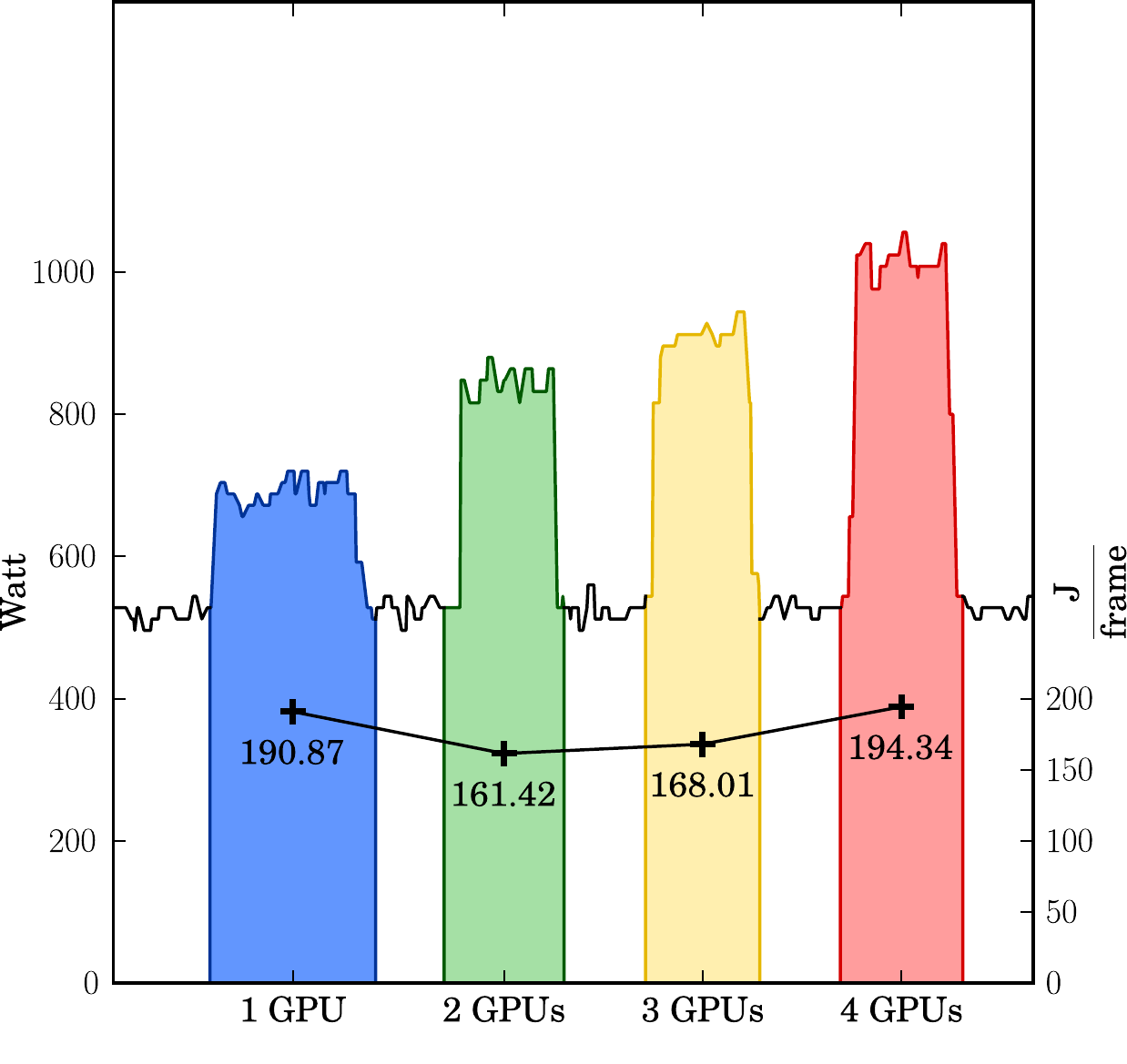}
\caption{Overall power drain and energy consumption per frame}
\label{fig:power}
\end{minipage}
\vspace{-0.37cm}
\end{figure}

The number of channels the algorithms operates on can be reduced using a principle component decomposition. Acceptable image quality is achievable by compressing the 32 data channels to no less than eight. Figure \ref{fig:speedup} shows the performance in frames per second as well as the speed-up for 2 different image resolutions, varying number of GPUs and different numbers of channels. Setup as well as pre- and post-processing steps are excluded from these measurements. Peak performance of $11.4$ frames per second is reached with the smaller matrix size, 8 channels and 4 GPUs. Due to unequal data distribution, the frame-rate is the same for 10 and 12 channels if 4 GPUs are used.

\begin{figure}[ht]
\vspace{-0.37cm}
\begin{minipage}[t]{0.48\linewidth}
\centering
\includegraphics[width=\linewidth]{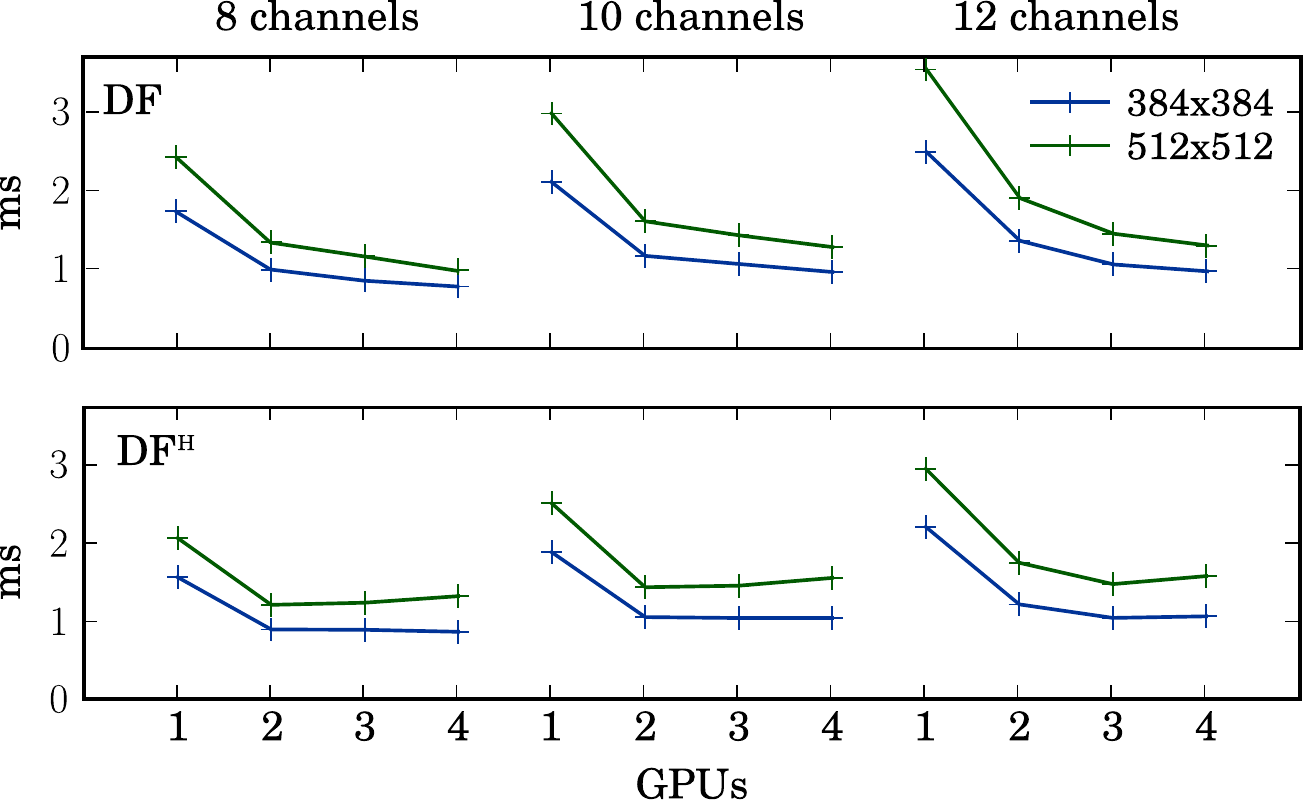}
\caption{$\FRDER$ and $\FRDER^{\text{H}}$ performance}
\label{fig:operators}
\end{minipage}
\hspace{0.1cm}
\begin{minipage}[t]{0.48\linewidth}
\centering
\includegraphics[width=0.98\linewidth]{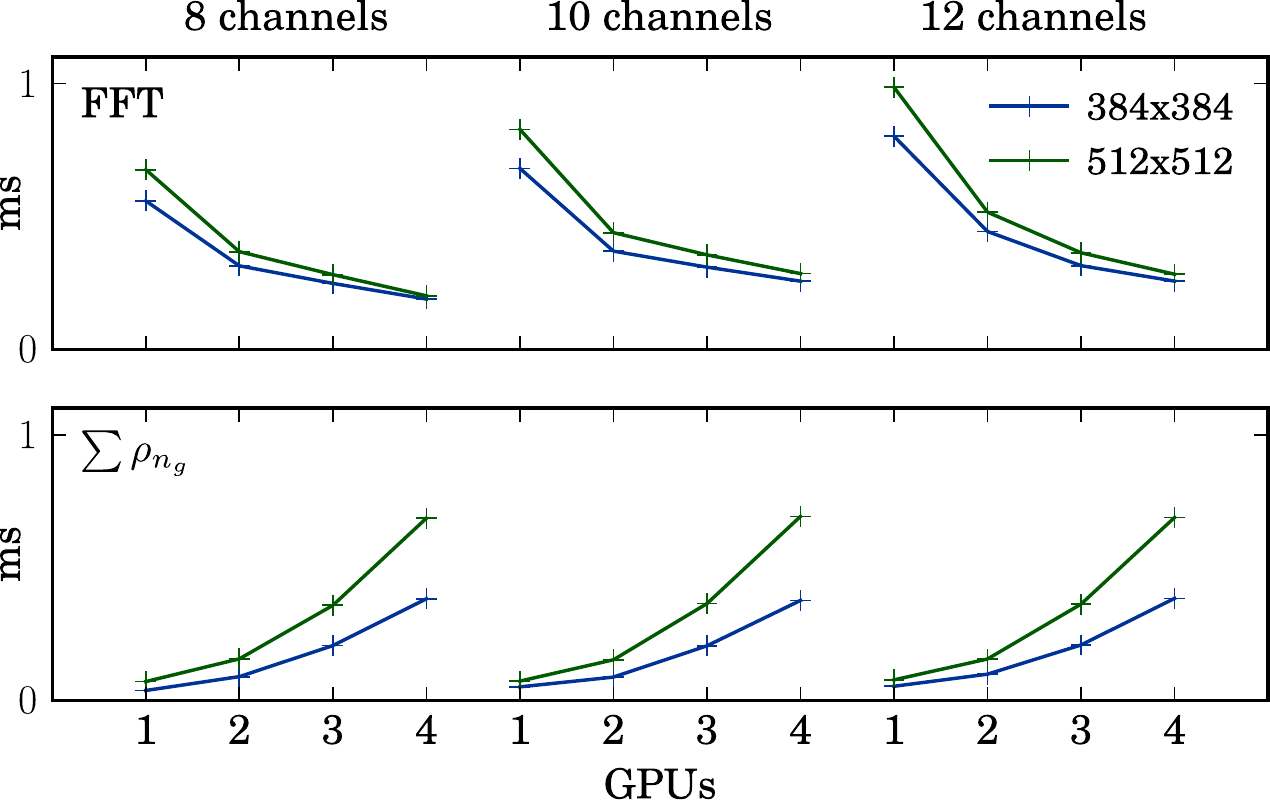}
\caption{FFT and all-reduce performance}
\label{fig:functions}
\end{minipage}
\vspace{-0.37cm}
\end{figure}

Figure \ref{fig:operators} shows a breakdown of the runtime for the 2 main operators $\FRDER$ and $\FRDER^{\text{H}}$. For the $\FRDER$ curve, the gradient increases if more channels are calculated. If the GPUs are given more work it can be distributed more efficiently. The performance of the Fourier transform greatly influences the performance of the overall image reconstruction algorithm as it is the most time-consuming operation. Its scalability is a function of batch- and matrix size as shown in figure \ref{fig:functions}. The operator $\FRDER^{\text{H}}$ includes peer-to-peer communication which causes a performance decrease if more than 2 GPUs are used. This communication overhead increases and neutralizes a large portion of the speed-up in this operator. Execution time even increases for 4 GPUs. Figure \ref{fig:functions} illustrates this interrelation. If only 1 GPU is used, the communication kernel effectively copies data on-device.

\begin{figure}[h]
\vspace{-0.37cm}
\centering
\includegraphics[width=\linewidth]{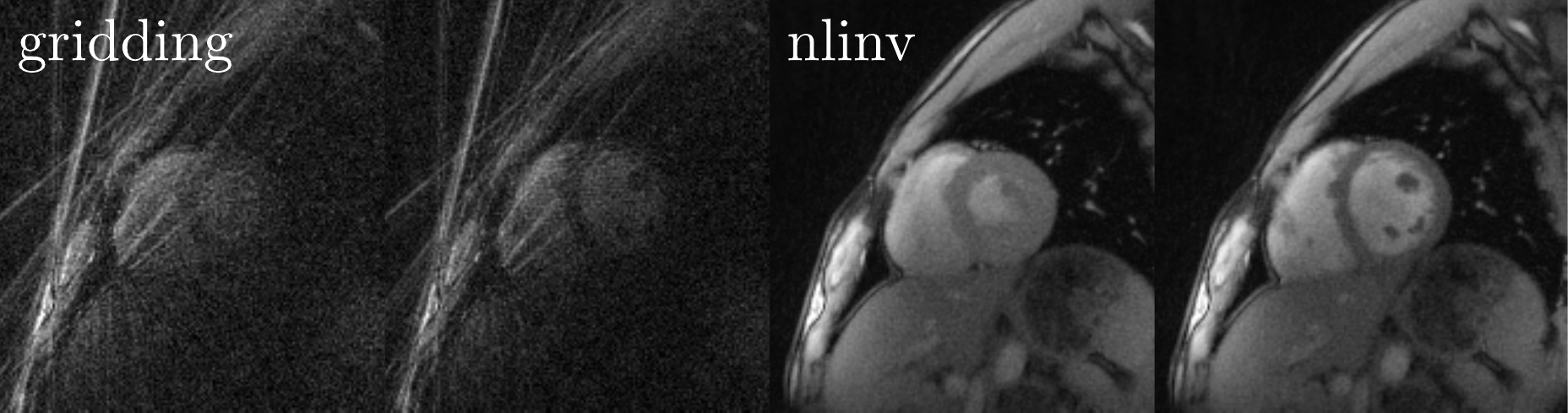}
\caption{Non-iterative (gridding) and non-linear inversion (nlinv) reconstruction of short-axis view of a healthy human heart at a temporal resolution of 33ms.}
\label{fig:comp}
\vspace{-0.37cm}
\end{figure}

Alongside the runtime performance of this algorithm the energy consumption of this multi-GPU implementation is investigated. We monitor power-drain of our system using its integrated baseboard management controller. Figure \ref{fig:power} shows the power consumption for reconstructing an MRI video with 400 frames using a variable number of compute-devices. The figure also shows the energy consumed per frame, calculated by integrating power over time. While using 2 and 3 GPUs is most power efficient, using 4 GPUs does not consume significantly more energy compared to single GPU image reconstruction.

Figure \ref{fig:comp} shows the difference between a non-iterative image reconstruction method and the non-linear inversion algorithm described in this work. While the extreme radial undersampling causes severe streaking artefacts, the iterative algorithm is capable of reconstruction high-quality images.

\section{Related Work}

Developers can choose between a number of tools and libraries such as CUDA, OpenCL and C++ AMP to develop code for single-GPU architectures. As to our knowledge, none exist that explicitly target single-node multi-GPU systems or provide facilities to describe algorithms regardless of the number of accelerators in the target system. There exist however a number of tools for GPU clusters such as the many GPUs package published by Barak et al. \cite{Barak2010} in 2010. The framework allows parallel OpenMP, C++ and unmodified OpenCL applications to run on multiple GPUs in a cluster. Network latency is a limiting factor in their implementation that is targeted at massive parallel and high performance computing applications. SkePU \cite{enmyren2010skepu} is a multi-backend (CUDA, OpenCL) library based on algorithmic skeletons \cite{cole1989algorithmic} for solving data-parallel problems on GPU and multi-GPU systems. The library can utilize multiple GPUs and automatically distributes data across all devices, the authors report issues with a multi-GPU gaussian blur implementation because of inter-GPU communication staged through main memory. A MGPU implementation of this algorithm could be designed to take the disjoint memory areas into account and utilize peer-to-peer memory transfer. AGILE \cite{Knoll2011} is an open source library for magnetic resonance image reconstruction using graphics card hardware acceleration. While restricted to a single GPU, their library targets the same applications as MGPU. Another high-level GPGPU library is Thrust \cite{hoberock2009thrust}: a selection of parallel algorithms closely following the C++ Standard Template Library. Thrust does not support data decomposition across multiple GPUs but can be used on multi-GPU systems through the use of the lower level CUDA API. ViennaCL \cite{rupp2010viennacl} is a high level linear algebra library that supports both GPUs and multi-core CPUs through OpenCL. While the library provides mechanisms to use multiple devices for computation, data decomposition and synchronization is left to the user.

The viability and usefulness of GPU and multi-GPU systems for real-time applications has previously been shown. A medical signal-processing application running on a multi-GPU system is described by Jang et al. \cite{jang2009multi}. Their implementation of an iterative image reconstruction algorithm for computed tomography achieves a speed-up of about $2$ when comparing single- and quad-GPU implementations. Verner et al. \cite{verner2011processing} present a hard real-time stream scheduling algorithm for GPU based systems and Chilingaryan et al. implement \cite{chilingaryan2011gpu} a near real-time GPU-base reconstruction software for synchrotron experiments. Suitability of GPUs for MRI reconstruction is shown early on by Hansen et al \cite{hansen2008cartesian}. Later Kim et al. \cite{kim2011high} compare implementations of an iterative algorithm for 3D compressive sensing MRI reconstruction on various many-core architectures including Nvidia GPUs and the novel Intel MIC architecture. A comprehensive discussion of multi-core and multi-GPU implementation of a novel iterative 3D algorithm for MRI with clinically feasible reconstruction time is presented by Murphy et al. \cite{6153065}.

\section{Conclusion}

We presented MGPU, a C++ template-based multi-GPU programming library for real-time applications. We demonstrated the benefits of the segmented container concept and pointed out the developer-friendly implementation of the various data transfer routines. An evaluation of our library through various micro-benchmarks yields expected results. Batched fast Fourier transforms and element wise matrix operations scale well across multiple devices. We have further demonstrated that the concepts of MGPU and its generic structure are suitable to extend existing single-GPU libraries with multi-GPU interfaces. High data locality due to low latency and high bandwidth, compact form factor compared to GPU clusters, manageable power consumption, low cost and the convenient but efficient tools MGPU provides are the reasons why multi-GPU systems are suitable to solve data-parallel real-time problems. MGPU is certainly not limited to the problem domain described in this work and we can imagine that general high performance computing problems can benefit from the combination of multi-GPU hardware and the MGPU framework.

The modular architecture of MGPU renders the extension of the current device interface with OpenCL support possible, which we plan to implement as a next step. This will enable us to compare the performance of Nvidia and AMD systems. Beyond that we will incorporate even more single-GPU libraries in our library. In addition, we plan to investigate how MGPU concepts can apply to emerging multi-core architectures such as Intel MIC and AMD Fusion.

Encouraged by our micro-benchmark results we used MGPU to extend an existing numerical algorithm for MRI image reconstruction and submit evidence that indeed, multi-GPU systems are suitable for speeding up signal-processing and real-time applications. We measured a speed-up by $1.7$ when using two compute-devices and a speed-up of more than 2 when using 4 GPUs. To overcome the problem of inter-GPU communication overhead when using 4 GPUs we plan to investigate alternative data decomposition schemes. Since the performance of the Fourier transform is the major determining factor of this algorithm we are also experimenting with different implementations that are capable of exploiting the sparsity of our data. 

Although the reconstruction frame rate does not yet match the temporal resolution of data acquisition process, the multi-GPU implementation of the algorithm is fast enough so that simultaneous recording and reconstruction is feasible. Thus, results are immediately available for scientists and physicians to interpret. The existence of an on-line iterative algorithm will ease the adoption of real-time MRI in a clinical setting. We expect that the high temporal resolution that is made possible by these advanced iterative algorithms will give radiologists new insights and might ultimately result in a more accurate diagnosis for patients.

% bibliography
\bibliographystyle{splncs03}

\bibliography{lncs}  % sigproc.bib is the name of the Bibliography in this case

\begin{thebibliography}{10}
\providecommand{\url}[1]{\texttt{#1}}
\providecommand{\urlprefix}{URL }

\bibitem{austern2000segmented}
Austern, M.: {Segmented Iterators and Hierarchical Algorithms}. Generic
  Programming pp. 80--90 (2000)

\bibitem{bakushinsk2004}
Bakushinski{\u\i}, A., Kokurin, M.: {Iterative Methods for Approximate Solution
  of Inverse Problems}, vol. 577. Kluwer Academic Pub (2004)

\bibitem{Barak2010}
Barak, A., Ben-Nun, T., Levy, E., Shiloh, A.: {A package for OpenCL based
  heterogeneous computing on clusters with many GPU devices}. In: Cluster
  Computing Workshops and Posters (CLUSTER WORKSHOPS), 2010 IEEE International
  Conference on. pp. 1--7. IEEE (2010)

\bibitem{Bergstrom2011}
Bergstrom, L.: {Measuring NUMA effects with the STREAM benchmark}. CoRR
  abs/1103.3225 (2011)

\bibitem{block2007tv}
Block, K., Uecker, M., Frahm, J.: {Undersampled radial MRI with multiple coils.
  Iterative image reconstruction using a total variation constraint}. Magnetic
  Resonance in Medicine  57,  1086--1098 (2007)

\bibitem{chilingaryan2011gpu}
Chilingaryan, S., Mirone, A., Hammersley, A., Ferrero, C., Helfen, L., Kopmann,
  A., dos Santos~Rolo, T., Vagovic, P.: {A GPU-Based Architecture for Real-Time
  Data Assessment at Synchrotron Experiments}. Nuclear Science, IEEE
  Transactions on (99),  1--1 (2011)

\bibitem{cole1989algorithmic}
Cole, M.: {Algorithmic Skeletons: Structured Management of Parallel
  Computation}. Pitman (1989)

\bibitem{dawes2009boost}
Dawes, B., Abrahams, D., Rivera, R.: {Boost C++ libraries}.
  http://www.boost.org

\bibitem{enmyren2010skepu}
Enmyren, J., Kessler, C.: {SkePU: A Multi-Backend Skeleton Programming Library
  for Multi-GPU Systems}. In: Proceedings of the fourth international workshop
  on High-level parallel programming and applications. pp. 5--14. ACM (2010)

\bibitem{hansen2008cartesian}
Hansen, M., Atkinson, D., Sorensen, T.: {Cartesian SENSE and k-t SENSE
  reconstruction using commodity graphics hardware}. Magnetic Resonance in
  Medicine  59(3),  463--468 (2008)

\bibitem{hoberock2009thrust}
Hoberock, J., Bell, N.: {Thrust: C++ Template Library for CUDA} (2009)

\bibitem{Huang2007compression}
Huang, F., Vijayakumar, S., Li, Y., Hertel, S., Duensing, G.: {A software
  channel compression technique for faster reconstruction with many channels}.
  Magnetic Resonance Imaging  26,  133--141 (2007)

\bibitem{jang2009multi}
Jang, B., Kaeli, D., Do, S., Pien, H.: {Multi GPU Implementation of Iterative
  Tomographic Reconstruction Algorithms}. In: Biomedical Imaging: From Nano to
  Macro, 2009. ISBI'09. IEEE International Symposium on. pp. 185--188. IEEE
  (2009)

\bibitem{jung2009k}
Jung, H., Sung, K., Nayak, K., Kim, E., Ye, J.: k-t focuss: A general
  compressed sensing framework for high resolution dynamic mri. Magnetic
  Resonance in Medicine  61(1),  103--116 (2009)

\bibitem{kim2011high}
Kim, D., Trzasko, J., Smelyanskiy, M., Haider, C., Dubey, P., Manduca, A.:
  {High-performance 3D compressive sensing MRI reconstruction using many-core
  architectures}. Journal of Biomedical Imaging  2011, ~2 (2011)

\bibitem{Knoll2011}
Knoll, F., Freiberger, M., Bredies, K., Stollberger, R.: {AGILE: An open source
  library for image reconstruction using graphics card hardware acceleration}.
  In: Proc. Intl. Soc. Mag. Reson. Med. 19:2554 (2011)

\bibitem{lustig2007sparse}
Lustig, M., Donoho, D., Pauly, J.: {Sparse MRI: The application of compressed
  sensing for rapid MR imaging}. Magnetic Resonance in Medicine  58,
  1182--1195 (2007)

\bibitem{6153065}
Murphy, M., Alley, M., Demmel, J., Keutzer, K., Vasanawala, S., Lustig, M.:
  {Fast $\ell_1$-SPIRiT Compressed Sensing Parallel Imaging MRI: Scalable
  Parallel Implementation and Clinically Feasible Runtime}. Medical Imaging,
  IEEE Transactions on  PP(99), ~1 (2012)

\bibitem{rupp2010viennacl}
Rupp, K., Rudolf, F., Weinbub, J.: {ViennaCL - A High Level Linear Algebra
  Library for GPUs and Multi-Core CPUs}. Proc. GPUScA pp. 51--56 (2010)

\bibitem{tsao2003k}
Tsao, J., Boesiger, P., Pruessmann, K.: k-t blast and k-t sense: Dynamic mri
  with high frame rate exploiting spatiotemporal correlations. Magnetic
  Resonance in Medicine  50(5),  1031--1042 (2003)

\bibitem{uecker2008image}
Uecker, M., Hohage, T., Block, K., Frahm, J.: {Image Reconstruction by
  Regularized Nonlinear Inversion - Joint Estimation of Coil Sensitivities and
  Image Content}. Magnetic Resonance in Medicine  60(3),  674--682 (2008)

\bibitem{uecker2010gpu}
Uecker, M., Zhang, S., Frahm, J.: {Nonlinear inverse reconstruction for
  real-time MRI of the human heart using undersampled radial FLASH}. Magnetic
  Resonance in Medicine  63,  1456--1462 (2010)

\bibitem{uecker2010realtime}
Uecker, M., Zhang, S., Voit, D., Karaus, A., Merboldt, K.D., Frahm, J.:
  {Real-time MRI at a resolution of 20 ms}. NMR in Biomedicine  23,  986--994
  (2010)

\bibitem{verner2011processing}
Verner, U., Schuster, A., Silberstein, M.: {Processing Data Streams with Hard
  Real-time Constraints on Heterogeneous Systems}. In: Proceedings of the
  international conference on Supercomputing. pp. 120--129. ACM (2011)

\end{thebibliography}

\end{document}